\documentclass[runningheads,a4paper]{llncs}

\usepackage{geometry}
\newcommand\MyHead[2]{%
\multicolumn{1}{|l|}{\parbox{#1}{\centering #2}}}

\usepackage{amssymb}
\usepackage{amsmath}
\usepackage{xspace}
\usepackage{graphicx}
\usepackage{dsfont}
\usepackage{algorithm}
\usepackage{algorithmic}
\usepackage[caption=false]{subfig}
\usepackage{url}
\DeclareMathOperator*{\argmax}{\arg\!\max}

\urldef{\mailsa}\path|{rom5161,mom5590, acharya}@cse.psu.edu|
\urldef{\mailsb}\path|mposs@bx.psu.edu|

\newcommand{\kmer}{$k$-mer\xspace}
\newcommand{\kmers}{$k$-mers\xspace}

\begin{document}

\mainmatter  

\title{A frame-based representation of genomic sequences for removing errors and rare variant detection in NGS data}

\titlerunning{A frame-based representation for error detection}

\author{Raunaq Malhotra $^1$%
\thanks{Corresponding author}%
\and Manjari Mukhopadhyay ${^1}$\and Mary Poss${^2}$ \and\\
Raj Acharya${^1}$}
\authorrunning{Malhotra R, Mukhopadhyay M, Poss M, Acharya R}

\institute{$^1$School of Electrical Engineering and Computer Science, Th Pennsylvania State University, 
University Park, PA, 16802, USA\\
$^2$ Department of Biology, The Pennsylvania State University, 
University Park, PA, 16802, USA\\
\mailsa, \mailsb\\
}

%
%

\maketitle

\begin{abstract}
We propose a frame-based representation of \kmers for detecting sequencing errors and rare variants in next generation sequencing data obtained from populations of closely related genomes. Frames are sets of non-orthogonal basis functions, traditionally used in signal processing for noise removal. We define a frame for genomes and sequenced reads to consist of discrete spatial signals of every \kmer of a given size. We show that each \kmer in the sequenced data can be projected onto multiple frames and these projections are maximized for spatial signals corresponding to the \kmer's substrings. Our proposed classifier, MultiRes, is trained on the projections of \kmers as features used for marking \kmers as erroneous or true variations in the genome.
 We evaluate MultiRes on simulated and real viral population datasets and compare it to other error correction methods known in the literature. MultiRes has 4 to 500 times less false positives \kmer predictions compared to other methods, essential for accurate estimation of viral population diversity and their \textit{de-novo} assembly. It has high recall of the true \kmers, comparable to other error correction methods. MultiRes also has greater than 95\% recall for detecting single nucleotide polymorphisms (SNPs), fewer false positive SNPs, while detecting higher number of rare variants compared to other variant calling methods for viral populations. The software is freely available from the GitHub link \url{https://github.com/raunaq-m/MultiRes}.
\end{abstract}

\section{Introduction}
Error detection in next generation sequencing (NGS) data is an important task for the accurate interpretation of polymorphisms in a population of genomes. For example, detecting sequencing errors and their correction is one of the first pre-processing steps in the assembly of haplotypes in viral populations \cite{QuasiRecomb,shorah,vga,vphaser2,reviewpaper} and in single nucleotide polymorphisms (SNPs) detection for viral populations \cite{vphaser2,lofreq,shorah}. 

The fundamental premise of error correction is that a true sequence in the genome is sampled via multiple reads whereas errors in sample preparation and sequencing occur randomly. Thus the replicates in NGS data can be used to identify true sequences and sequencing errors occurring in small numbers of reads can be differentiated from the true sequences. A number of error correction methods for NGS data from haploid or diploid genomes (eg. \cite{quake,bless,bfc,musket,Hammer}) use this principle to identify error-free regions in the sampled reads, and then correct the regions with sequencing errors. 

However, error correction is especially challenging for datasets from a heterogeneous population of closely related genomes, such as those observed in viruses within a host \cite{reviewpaper,Niko}. Typically, haplotypes that exist at a low frequency in the population, also known as rare variants, are sampled less frequently and their prevalence is close to that of sequencing errors in the NGS data. While the reasons for viral diversity are well understood \cite{domingo2012viral,domingo,nowak}, a mutation error rate close to the sequencing error rate for many viral polymerases makes it difficult to detect the full mutation spectrum of viruses. Additionally, preparation of RNA viruses such as the Hepatitis C Virus (HCV) and the Human Immunodeficiency virus (HIV) for sequencing include a reverse-transcriptase (RT) step and a Polymerase Chain Reaction (PCR) step, both of which introduces errors in the sequenced reads.

Error correction for viral populations typically involves aligning reads to a known reference sequence, and then marking the bases observed corresponding to every position in the genome as either as a true variant or a sequencing error based on a probabilistic model followed by hypothesis testing \cite{shorah,QuasiRecomb,HaploClique,vphaser2,lofreq}. However, the possibility of extensive diversity in a virus population hinders the accurate mapping of reads to any one reference.

Alternatively, sampled reads are broken into small fixed length sub-strings called \kmers and their counts are used for error correction. The \kmer counts are modeled using a generative model to determine a \kmer frequency threshold \cite{kec,quake,bless,bfc}; \kmers with counts greater than the threshold are marked as true \kmers and used for correcting other \kmers. The size of the \kmer and the frequency threshold are important parameters affecting an algorithm's performance. For viral populations, a single threshold is not suitable as haplotypes occur at different relative frequencies. Additionally, the choice of \kmer size either decreases the evidence for a segment of the genome for a large $k$, or combines evidences from multiple segments for a small $k$. Moreover, most error correction algorithms based on \kmer frequencies \cite{bfc,bless,quake,musket} have been designed assuming the reads are sampled from a single diploid genome; thus these models are not applicable to single stranded viral populations.  

Currently, a number of time and memory efficient \kmer counting algorithms are available \cite{dsk,kmc2}. Thus, choosing an appropriate size of \kmer is possible by performing \kmer counts at multiple sizes \cite{kmergenie}. However, a single appropriate \kmer size for error correction in viral populations is restrictive in nature, as a combination of different sized overlapping \kmers, although redundant, can provide richer information. 

The notion of utilizing redundant information from different \kmer lengths simultaneously has parallels in signal processing, where signals analysis at different resolutions has traditionally been used for noise reduction \cite{frameserror}. Signals are represented as projections over a frame, which consists of a series of non-orthonormal basis functions \cite{duffin,daubechies1,daubechies2}; these projections are used for error removal and signal recovery \cite{image1,ronshen1}.

In this paper, we propose frames for genomic sequences and define genomic sequences as spatial signals. The spatial signals for the genomic sequences, the sequenced reads and \kmers in the NGS data are projected on the frames and these projections are used for removing sequencing errors and detecting rare variants in the NGS data. We propose a Random Forest Classifier, MultiRes, trained on the frame-based projections of NGS data for the detection of rare variants and the removal of sequencing errors. 

The frames are defined such that each frame corresponds to \kmers of a given size. The classifier in MultiRes is trained on a simulated dataset that models NGS data generated from a replicating viral population. MultiRes can then by used for error detection on real sequencing data obtained from the same sequencing technology.

We evaluate the performance of MultiRes on simulated and real datasets, and compare it to other error correction methods BLESS \cite{bless}, Quake \cite{quake}, BFC \cite{bfc}, and Musket \cite{musket}. We also compare our results to BayesHammer \cite{bayeshammer} and Seecer \cite{seecer}, which can handle variable sequencing coverage across the genome and polymorphisms in the RNA sequencing data respectively. MultiRes has a high recall of the true \kmers, comparable to other methods and has 5 to 500 times better removal of erroneous \kmers compared to other methods.

We also compare SNP detection from MultiRes to variant calling methods VPhaser-2 \cite{vphaser2}, LoFreq \cite{lofreq} and haplotype reconstruction method ShoRAH \cite{shorah}. MultiRes has the higher recall of true SNPs compared to the SNPs called by VPhaser-2, LoFreq and ShoRAH on both simulated and real datasets, and misses the least number of true SNPs amongst all methods. 

MultiRes has applications for studying the large scale variation in closely related genomes, such as viral populations. The complexity of De Bruijn graphs, useful for studying structural variants and rearrangements in the population, increases because of sequencing errors. Our method can provide a compact set of \kmers while still retaining high recall of the true \kmers, which can be utilized for constructing the graph. Additionally, the error corrected \kmers predicted by MultiRes can be directly used for understanding the SNPs observed in the viral population to a high degree of accuracy. 

 
\section{Methods}

\subsection{Definitions}
\subsubsection{Definition of Frames} 
A family of functions $\{f_{m}\}_{m\in \mathds{N}}$ in a Hilbert Space $\mathcal{H}$ is a \textbf{frame} for $\mathcal{H}$ if there exists positive constants $\alpha$ and $\beta$ such that 
\begin{equation}
\label{framedefn}
\alpha \cdot \|f\|^{2} \leq  \sum_{m}{ |\langle f, f_m \rangle|^{2}} \leq \beta \cdot \|f\|^{2}
\end{equation}
for all $f\in \mathcal{H}$ (\cite{duffin,kaiser2010friendly,daubechies2,daubechies1}). Here, $|\langle f, f_m \rangle|$ is the inner product of the function $f$ with a function $f_m$ in the family, while $\|f\|^{2}$ is the inner product of the function $f$ with itself in the space $\mathcal{H}$, also known as the norm of $f$. The functions in $\{f_{m}\}_{m\in \mathds{N}}$ are not necessarily orthogonal to each other. Thus, $|\langle f, f_m \rangle|$ or the projections form a redundant representation of the function $f$ on the family $\{f_m\}_{m\in \mathds{N}}$. Redundant projections of a signal have been exploited for noise reduction in signal processing \cite{frameserror}. 

\subsubsection{Genomes as Spatial Signals}
The genomic sequence of a haplotype in a viral population can be represented as a discrete spatial signal with a non-zero signal at a spatial position representing the base observed at the corresponding genomic position. Notationally, a haplotype $H$ of length $|H|$ is represented as a discrete 4-dimensional spatial signal $\{H_n\}$ where the signal at the $n^{th}$ position is defined by a 4-dimensional vector $H[n] = (x_{A}[n],x_{C}[n],x_{G}[n],x_{T}[n])^{\prime}$ for $n = \{0,1,2,\ldots,|H|-1\}$. The element $x_{b}[n]$ corresponds to the base $b \in \{ A,C,G,T\}$ observed at position $n$ of the genome. A haplotype's spatial signal is represented by $\{ H_n\}$, or $H$ in short, and $H[n]$ is used to denote its $n^{th}$ sample in the rest of the paper. 

For the spatial signal $\{H_n\}$ of a single haplotype two conditions hold true at each position $n$: (i) $x_{b}[n] \in \{0,1\}$, and (ii) $\sum_{b}{x_{b}[n]} = 1$ for $b \in \{A,C,G,T\}$. In other words, the non-zero entry in the 4-dimensional vector $(x_{A}[n],x_{C}[n],x_{G}[n],x_{T}[n])^{\prime}$ at position $n$ indicates that the corresponding base is observed at that position.

For a viral population $\mathbf{H}$ containing a collection of viral haplotypes, $\mathbf{H} =  \{H_1,H_2, \ldots, H_P\}$, we can define a spatial signal for each of the individual haplotypes $\{(H_{i})_{n}\}$ for $i=\{1,2,3,\ldots,P\}$, and $n$ denotes the spatial position in the signal for the haplotype $H_i$, as mentioned above. 

For a collection of reads $\mathbf{R}=\{R_1,R_2,\ldots,R_N\}$ sampled from the viral population $\mathbf{H}$, each read can be represented by a spatial signal $\{(R_i)_{n} \}$. The signal is non-zero for a fixed number of spatial positions equal to the read length, and is defined as described above. 


\subsubsection{Translation Operator for Spatial Signals} A translation operator $T$ shifts a genomic signal $H$ by one spatial position to the right. Mathematically, the translated signal $\{(T\cdot H)_n \}$ is defined as
${(T\cdot H)}[n] = H[n-1]$
for its $n^{th}$ sample.

A genomic signal translated by $r$ bases to the right is denoted as $\{(T^r\cdot H)_n \}$, where $(T^r\cdot H)[n] = H[n-r]$. 

The translation operator is useful as an identical subsequence of bases (or a \kmer) at multiple locations in the genome can be represented as translations of a single spatial signal. 

\subsubsection{Spatial Signal for Sampled Reads}
A read sampled from a specific location in the genome can be represented by translation of its spatial signal by an appropriate number of bases. Thus, a single spatial signal $\{ \mathbf{R}_n\}$ can be constructed for all the reads where the signal at the $n^{th}$ position represents the distribution of bases observed at that position in all the haplotypes. Here $\mathbf{R}[n] =  (r_{A}[n],r_{C}[n],r_{G}[n],r_{T}[n])^{\prime}$ where $r_{A}[n]$ denotes the number of aligned reads which have the base $A$ at position $n$, and so on.

\subsubsection{Inner Products of Spatial Signals} 
As the genomes are represented as 4-dimensional spatial signals, the standard definitions of vector addition and scalar multiplication apply to them. For two genomic signals $X$ and $Y$ defined above, the inner product of the two signals is defined as :
$\langle X, Y \rangle = \sum_{n} {\sum_{b \in \{A,C,G,T\}} {x_{b}[n]\cdot y_{b}[n]}}$
The inner product of two genomic signals measures the similarity of the two sequences represented by $X$ and $Y$. When $X$, $Y$ represent genomic signals of a single haplotype, the inner product reduces to a Kronecker delta product between the elements of $X$ and $Y$ at each spatial location $n$. 

If $X$ is a spatial signal from the sequenced reads and $Y$ represents a haplotype $H_i$, then the inner product provides a measure of the concordance between the sequenced reads and the haplotype $H_i$. 


\subsection{Frames for Genomic Signals} 
\subsubsection{Representation of \kmers of a Given Size as a Set of Signals}
A set $\mathbf{C_k}$ of signals is defined as a collection of $4^k$ discrete 4-dimensional spatial signals. The signals in the set $\mathbf{C_k}$ have a one-to-one correspondence to the set of all possible $4^k$ \kmers. 

Signals in the set $\mathbf{C_k}$ are denoted as $C_{k,l}$, where the index $k$ denotes that the signal belongs to the set $\mathbf{C_k}$ and the index $l\in \{1,2, \ldots, 4^k\}$ denotes one of the $4^k$ \kmers. Each signal $C_{k,l}$ is non-zero only at spatial positions $[0,(k-1)]$ and has exactly one non-zero entry in the four axes at a given spatial position. 

For example, the set $\mathbf{C_1}$ consists of four spatial signals: $\mathbf{C_1} \equiv (\{(C_{1,A})_n\},\{ (C_{1,C})_n\},\{(C_{1,G})_n\}, \{(C_{1,T})_n\})$, where $\{(C_{1,A})_n\}$ is defined as follows: 

\begin{equation}
C_{1,A}[n] = 
\begin{cases}
(1,0,0,0) & \text{if } n=0 \\ 
(0,0,0,0) &  \text{otherwise }
\end{cases}
\end{equation} 

\noindent The signals $\{ (C_{1,C})_n\}$, $\{(C_{1,G})_n\}$, and $\{(C_{1,T})_n\}$ are defined in similar fashions where, at $n=0$, the second, third, and fourth dimension of the 4-dimensional vector is respectively one. Thus, the signals $\{(C_{1,A})_n\}$, $\{ (C_{1,C})_n\}$, $\{(C_{1,G})_n\}$, and $\{(C_{1,T})_n\}$ correspond to the $1$-mers (A,C,G,T) being observed at the first position of the genome.  
 
%
%

A pictorial representation for a signal in the set $\mathbf{C_7}$ corresponding to $7$-mer AATCGAT shows that the $7$-mer can be trivially reconstructed from the signal by replacing the non-zero base value on the y-axis at the spatial position depicted by the x-axis (Figure \ref{fig:Timings}). 


\subsubsection{Representation for $\mathbf{C_{k}}$ from $\mathbf{C_{k-1}}$}
We can also obtain the set $\mathbf{C_k}$ of signals iteratively using the signals in the set $\mathbf{C_{k-1}}$ and the translations of the signals in the set $\mathbf{C_1}$. 

\begin{multline}
\mathbf{C_k} \equiv \{ C_{(k-1),l} + (T^{(k-1)}\cdot C_{1,b}) :  \quad l \in \{A,C,G,T\}^{(k-1)}, 
 \quad b \in (A,C,G,T) \}
\end{multline}

\noindent where $C_{(k-1),l}$ is a signal in the set $\mathbf{C_{k-1}}$ and $(T^{(k-1)}\cdot C_{1,b})$ is the translation of the signal $C_{1,b}$ to the position $(k-1)$. In other words, the signal for $k$-bases in the genome can be obtained by taking a signal for the first $(k-1)$-bases and appending it with the translated version of a signal in the set $\mathbf{C_1}$. The set of signals in $\mathbf{C_k}$ thus denote the $4^k$ possible \kmers that exist in $\{A,C,G,T\}^{k}$.

\subsubsection{Family of Signals from set $\mathbf{C_k}$}
A family of signals is obtained by translating the signals in the set $\mathbf{C_k}$ to any spatial position $r$.  Mathematically, 
$\mathbf{C_{k,\mathds{N}}} \equiv \{ T^r \cdot C_{k,l} : r \in \mathds{N}, l \in \{A,C,G,T\}^{k} \}$
 which corresponds to all translations of the signals in the set $\mathbf{C_k}$ by $r \in \mathds{N}$ spatial positions. The family is indexed by three parameters: $k$ indicating the set $\mathbf{C_k}$ of signals used, $l$ indicating a particular signal of the possible $4^k$ signals in the set $\mathbf{C_k}$, and $r$ indicating a spatial position to which the signal $C_{k,l}$ is translated. 

The family of signals $\mathbf{C_{k,\mathds{N}}}$ physically corresponds to signals for \kmers observed at any position in a genome. For example, the signal $T^r\cdot C_{k,l}$ corresponds to the signal $C_{k,l}$ translated by $r$ bases to the right, so that it is non-zero between positions $r$ to $r+k-1$. In other words, it corresponds to the \kmer represented by $C_{k,l}$ starting at position $r$ in the genome. Thus, as $C_{k,l}$ varies over the set $\mathbf{C_k}$, the family $\mathbf{C_{k,\mathds{N}}}$ can represent all \kmers at all genomic positions. This is important as the genome of any haplotype can be represented in terms of a family of signals for a given size $k$. 

\subsubsection{Family of Generating Signals $\mathbf{C_{k,\mathds{N}}}$ as a Frame for Genomic Signals}
We define a function $\tilde{f_k} : H \rightarrow \mathds{N}$, given a genomic signal $H$ and a family $\mathbf{C_{k,\mathds{N}}}$, as: 

$\tilde{f_k} (l,r) \equiv \langle H, (T^r\cdot C_{k,l}) \rangle$
, $\tilde{f_k} (l,r)$ denotes the inner-product of the signal $H$ with the generating signal $C_{k,l}$ that has been translated by $r$ positions, a member of the family $\mathbf{C_{k,\mathds{N}}}$. The function $\tilde{f_k} (l,r)$ is indexed by $l$ and $r$, where $l$ ranges between $1$ to $4^k$ and $r$ ranges over the spatial positions. 

For example, consider the projections of $H$ onto the family $\mathbf{C_{1,\mathds{N}}}$. Here, $\tilde{f_1}(l,r)$ denotes the unique projections of $H$ onto the signals in the family $\mathbf{C_{1,\mathds{N}}}$. As a signal $T^r \cdot C_{1,l} \in \mathbf{C_{1,\mathds{N}}}$, is non-zero only at one spatial position, namely $r$, and the genomic signal $H$ has only non-zero entry at each spatial position, the function $\tilde{f_1}(l,r)$ is equal to one only in the direction of the base observed in $H$ at a position $r$. This allows us to uniquely express $H$ using its projections, namely:  
$H[n] = \sum_{b\in\{A,C,G,T\}} { \tilde{f_1}(l,r) \cdot (T^r\cdot C_{1,b})[n]}$.

This is easy to see, as the members of the family $\mathbf{C_{1,\mathds{N}}}$ are orthonormal and are non-zero at only one spatial position and for only one of the four dimensions. Each of the signal samples a single base at a single spatial position in the signal $H$ and the collection of projections $\tilde{f_1}(l,r)$ is another way to specify the genomic signal $H$. This is very similar to the concept of sifting property of discrete delta functions in signal processing. 

The family $\mathbf{C_{k,\mathds{N}}}$ forms a frame for representing genomic signals \cite{kaiser2010friendly} and can be used for representing any genomic signal $H$. In other words, genomic signal $H$ can be described using its projections onto the family of functions $\mathbf{C_{k,\mathds{N}}}$. These projections can be used to completely specify the signal $H$. 

In order to show that $\mathbf{C_{k,\mathds{N}}}$ forms a frame, we show that Equation \ref{framedefn} holds for all genomic signal $H$ with the inner-product as defined above. We show two properties for the function $\tilde{f_k}(l,r)$ that demonstrate $\mathbf{C_{k,\mathds{N}}}$ is indeed a frame for genomic signals and the fact that only analyzing the \kmers present in a genome is sufficient for its representation.

\begin{proposition}
\label{kmermaxvalue}
The maximum value of the projection $\tilde{f_k}(l,r)$ for signal $H$ from a single genome, $\max_{l}{\tilde{f_k}(l,r)} = k. $
\end{proposition}
%
%
%

As the projection $\tilde{f_k}(l,r)$ is bounded by $k$ for a single genome and there are only finite number of non-zero projections ($4^k \cdot |H|$ total projections) on the family $\mathbf{C_{k,\mathds{N}}}$, the upper bound $\beta$ for the frame definition (Equation \ref{framedefn}) can be obtained trivially $(\beta = k )$. Even for the sampled reads signal $\mathbf{R}_n$, its projections are again bounded by $k \cdot \max{ |\mathbf{R}[n]|}$ in a $k-$spatial domain window. The lower bound $\alpha$ can be obtained as all non-zero signals have projection of at least one. Thus, the bounds in the frame definition are still valid and the family of signals $\mathbf{C_{k,\mathds{N}}}$ is indeed a frame. 

We next show that the projection of a genomic signal $H$ is maximized when the generating signal is for the \kmer matching the genomic signal $H$.

\begin{proposition}
\label{kmermax}
The \kmer $u$ observed at position $r$ in the signal $H$ corresponds to the signal in the family $\mathbf{C_{k,\mathds{N}}}$ for which the inner product  $ \langle H, (T^r\cdot C_{k,l}) \rangle $ is maximized 
$ u = \argmax_{l} \tilde{f_k}(l,r) $.
\end{proposition}


Proposition \ref{kmermax} suggests that of all the projections of $H$ onto the family $\mathbf{C_{k,\mathds{N}}}$ in a $k-$spatial positions window, the maximum projection is on the generating signal corresponding to the \kmer of $H$ in the same window. We refer to this projection as the \textit{maximal projection} for a signal $H$ in a $k-$base window.  This is helpful as it gives the mathematical proof from the signals domain about something that is well known in practice, namely, that the a genome can be expressed in terms in terms of \kmers of a given length. 

As the family $\mathbf{C_{k,\mathds{N}}}$ forms a frame for the discrete four-dimensional spatial signals, the projections of the reads signal onto this family forms can be used to represent the signal.

\subsubsection{Representation of Sampled Reads Signal $\{ \mathbf{R_n} \}$ as Projections on Frames}
\label{sampledreads}
The maximal projections of the sampled reads signal $\{\mathbf{R}_n\}$ onto the frame $\mathbf{C_{k,\mathds{N}}}$ can be computed using the above definitions and we show below that they are directly proportional to the count of the \kmers present in the sampled reads. Consider the projection of the reads signal $\{\mathbf{R}_n\}$ onto a signal $T^r \cdot C_{k,l} \in \mathbf{C_{k,\mathds{N}}} $: 
$\langle \mathbf{R}[n], (T^r \cdot C_{k,l}) \rangle = \sum_{i} { \langle R_i[n], (T^r \cdot C_{k,l} ) \rangle } $
 where the summation on the right is over all the reads. As the signal $ (T^r \cdot C_{k,l} ) $ is non-zero only for spatial locations $r$ to $(r+k-1)$, only reads sampled from that segment in the viral genomes would contribute terms to the summation. If we only focus on the maximal projections of the reads signal, the individual inner-products for a read $R_i[n]$ will attain their maximum value $k$  (using propositions \ref{kmermaxvalue} and \ref{kmermax}) when the generating signal $C_{k,l}$ matches the \kmer present in $R_i[n]$ in this window. Thus, for a particular generating signal $T^r \cdot C_{k,l}$, the number of times it achieves its maximum will be exactly equal to the number of times a \kmer is observed in the all the reads at the positions $r$ to $r+k-1$. 

If one considers all the maximal projections of the reads signal $\{ \mathbf{R}_n \}$ onto the frame $\mathbf{C_{k,\mathds{N}}}$, the values of the projections are in fact equal to the counts of \kmers times the constant $k$. This is crucial as the properties of projections of signals can now be applied to the counts of \kmers for distinguishing between erroneous and rare variant \kmers, both of which have low \kmer counts in the sampled reads. 

The above equation implies that as long as the size of $k$ is large enough that a \kmer can only be sampled from a single location in the genome, all its observed counts would contribute to exactly one generating signal $T^r \cdot C_{k,l}$ in the family $\mathbf{C_{k,\mathds{N}}}$. In viral populations, where repeats are small, it is possible to choose reasonable values of $k$ for the above to be true. Thus, the choice of $k$ for the frame $\mathbf{C_{k,\mathds{N}}}$ is important and should be large enough such that a \kmer only occurs once in the haplotypes. On the other hand, it should be smaller than the read lengths so that \kmer counting is still meaningful. 

The minimum $k$ can be approximated by ensuring that the probability of picking a string of length $|H|$ where all \kmers in it occur only once \cite{quake}. Thus the probability of picking approximately $|H|$ unique \kmers out of a set of $4^{k/2}$ (considering reverse complements) should be low. We set $ 2 \cdot |H| / 4^k \approx \epsilon $, where $\epsilon$ is a small number, to determine the smallest possible choice of $k$ ($k_{min}$) for the frame $\mathbf{C_{k_{min},\mathds{N}}}$.


\subsubsection{Representation of \kmers as a Series of Frames for Error Detection} 
The reads signal can be projected onto multiple families of signals as additional redundancy helps in reducing the noise in the signal, as long as the noise is random \cite{mallat1999wavelet}. If the parameter $k$ for a frame $\mathbf{C_{k,\mathds{N}}}$ is greater than $k_{min}$, the maximal projections of the reads signal will be proportional to the counts of the \kmers, and one can choose multiple such frames for representing the reads. For example, given the frames $\mathbf{C_{k,\mathds{N}}}$, $\mathbf{C_{k^{\prime},\mathds{N}}}$, $\mathbf{C_{k^{\prime\prime},\mathds{N}}}$ for $k>k^{\prime}>k^{\prime\prime}> k_{min}$, the maximal projections of the reads signal on these three frames will correspond to the counts of the \kmers, $k^{\prime}$-mers, $k^{\prime\prime}$-mers of all the reads. 

The projections of the reads signals onto a series of frames can be used for detection of erroneous windows and of rare variants similar to noise removal in signal processing \cite{image1,ronshen1}. As the projections of the reads are obtained in a window of fixed size we also perform detection of rare variants and errors based on the \kmers and their projections onto a series of frames. 

For a \kmer $u$ occurring $c(u)$ times in the reads we denote its spatial signal as $\{u_n\}$. The maximal projection of $\{u_n\}$ onto the frame $\mathbf{C_{k,\mathds{N}}}$ is $k\cdot c(u)$. For $k$-values $(k^{\prime}, k^{\prime\prime}$  in the range $[k_{min},k]$, the signal $\{ u_n \}$ can also be projected onto the frames $\mathbf{C_{k^{\prime},\mathds{N}}}$ and $\mathbf{C_{k^{\prime\prime},\mathds{N}}}$. As before, the maximal projections of $\{ u_n \}$ onto these frames are equal to the counts of $k^{\prime}$-mers and $k^{\prime\prime}$-mers present within $u$ in their respective dimensions and frames. 

\subsection{Classification of \kmers using Maximal Projections}
\subsubsection{Classes for \kmers classification}
The counts of the \kmers along with the counts of their sub-sequences (sub \kmers within a \kmer) are used as features for training a classifier. The true \kmers observed in the viral haplotypes with counts in the reads less than a threshold $T_{High}$ are defined to be rare variant \kmers, while the rest of \kmers with counts less than $T_{High}$ are erroneous \kmers. The \kmers that occur at counts greater than $T_{High}$ are known as common \kmers, as they occur frequently in the viral haplotypes. The common \kmers are assumed to be error-free and the classifier is trained only for the erroneous and rare variant \kmers.

\subsubsection{MultiRes: Classification Algorithm for Detecting Sequencing Errors and Rare Variants}
We define a classifier, $EC$, for classifying a \kmer as erroneous, a rare variant, or a common \kmer in the dataset. Algorithm \ref{alg:01} describes MultiRes, the proposed algorithm for detecting rare variants and removing sequencing errors. The algorithm takes as input the sampled reads $\mathbf{R}$, the classifier $EC$, an ordered array $(k,k^{\prime},k^{\prime\prime})$, and a threshold parameter $T_{High}$. It outputs for every \kmer observed in the sampled reads a status: whether the \kmer is erroneous or a rare variant. 

It first computes the counts of \kmers, $k^\prime$-mers, and $k^{\prime\prime}$-mers using the dsk \kmer counting software \cite{dsk}. The \kmers $u$ that have counts greater than $T_{High}$ are marked as true \kmers while the rest of the \kmers are classified using the classifier $EC$ based on their projections onto the two frames. 

The classifier $EC$ captures the profile of erroneous versus rare variant \kmers from Illumina sequencing of viral populations. We used the software dsk (version 1.6066) \cite{dsk} for \kmer counting, which can perform the \kmer counts in a limited memory and disk space machine quickly. The run time of MultiRes is linearly dependent on the number of unique \kmers in a dataset, as once the classifier $EC$ is trained, it can be used for all datasets, and it can be easily parallelized. 

\begin{algorithm}[h]
\caption{\label{alg:01} MultiRes: Error Correction of the sampled reads $\mathbf{R}$ by frame-based classification of \kmers }
\textbf{Input:} Sampled reads $\mathbf{R}$, classifier $EC$, $(k,k^{\prime},k^{\prime\prime})$, $T_{High}$ \\
\textbf{Output:} $ y = EC(u)$ for all \kmers $u$ in the sampled reads, where $y = \{ 0,1\} $ (True \kmer, Erroneous \kmer)
\begin{algorithmic}[1]
\STATE Compute counts $c(\cdot)$ of \kmers, $k^{\prime}$-mers and $k^{\prime \prime}$-mers from the sampled reads $\mathbf{R}$. 
\FOR { each \kmer $u=(u_1,u_2,\ldots,u_k)$}
\IF{$c(u) < T_{high}$}
	\STATE $ F(u) = [c(u), c(u_1,u_2,\ldots,u_{k^{\prime}}), $
	$c(u_2,u_3,\ldots,u_{k^{\prime}+1}), \ldots, c(u_{k-k^{\prime}+1},\ldots,u_{k}), $
 	$ c(u_1,u_2,\ldots,u_{k^{\prime \prime}}), $
 	$c(u_2,u_3,\ldots,u_{k^{\prime \prime}+1}),$
 	$ \ldots, c(u_{k-k^{\prime \prime}+1},\ldots,u_{k})] $
	\STATE  Use classifier $EC$ to classify the \kmer $u$ using its features $F(u)$ 
	\STATE Output $\{ 0, 1 \}$ for \kmer $u$ based on the classifier output
\ELSE
\STATE Output $\{1\}$ for the \kmer $u$ // Mark the \kmer as true
\ENDIF
\ENDFOR
\end{algorithmic}
\end{algorithm}

\subsubsection{Simulated Data for Classifier Training}
MultiRes assumes the availability of a classifier $EC$ which can distinguish between the erroneous and rare variant \kmers based on their projections onto frames. We use simulated datasets to train a series of classifiers and set $EC$ to the classifier which has the highest accuracy. A simulated a viral population consisting of 11 haplotypes is generated by mutating 10\% of positions on a known HIV-1 reference sequence of length 9.18 kb (NC\_001802). The mutations introduced are randomly and uniformly distributed so that the classifier is not biased towards the distribution of true variants. Thus, we introduces a total of 195K ground truth unique $35$-mers in the simulated HIV-1 dataset. 

We next simulate Illumina paired-end sequencing reads using the software dwgsim (\url{https://github.com/nh13/DWGSIM}) at 400x sequencing coverage from this viral population. The status of each \kmer in this dataset is known as being erroneous, rare variant or a common \kmer. A 10\% random sub-sample of all the \kmers is used as training dataset. 

In order to train a classifier, we need to choose the size of the \kmer, the sizes of $k^{\prime}$-mers for computing the projections of \kmer signals, and the number of such projections needed. The choice of the smallest of $\{k,k^{\prime},\ldots\}$ should be above the minimum length $k_{min}$ to ensure that each \kmer still corresponds to a unique location on a viral genome. 

For HIV populations, with genome length 9180 base pairs (9.1 kbp) and taking $\epsilon = 0.001$ (a small value, as mentioned before), the minimum length of \kmer is $k_{min} = \lceil \log_{4} {2\cdot G / \epsilon} \rceil = \lceil 12.06 \rceil = 13  $. As in signal processing domain, we choose $k \approx 3\cdot k_{min} = 35$ (an integral multiple of $k_{min}$) as the largest \kmer, and consider its projections on frames of sizes ranging from $13$ to $35$. 

MultiRes assumes that \kmers above the threshold count $T_{high}$ are error-free, and only classifies the \kmers with counts less than $T_{high}$. The choice of $T_{high}$ should ensure that the probability of erroneous \kmers with counts above $T_{high}$ is negligible. We use the gamma distribution model mentioned in the Quake error correction paper \cite{quake} for modeling erroneous \kmers, as it nicely approximates the observed distribution of errors. Based on this gamma distribution, we set $T_{high}=30$ for the simulated HIV population data. The classifiers are therefore trained on $35$-mers with counts less than 30. 

Three training datasets consisting of both erroneous and rare variant $35$-mers are generated. The features in the three datasets are the projection of the $35$-mers onto (i) the frame $\{\mathbf{C_{23,\mathds{N}}}\}$, (ii) the frame $\{ \mathbf{C_{13,\mathds{N}}}\}$, and (iii) a combination of both frames, namely $\{ \mathbf{C_{23,\mathds{N}}}, \mathbf{C_{13,\mathds{N}}}\}$ (Figure \ref{fig:02} a). Using Proposition \ref{kmermax}, the features translate to the counts of the $13$-mers and $23$-mers observed within the $35$-mer along with the counts of the $35$-mer. We observed $11.9$ million unique $35$-mers in the simulated HIV-1 population, from which features from 76000 erroneous $35$-mers and 32000 true variant $35$-mers distributed uniformly over counts $1$-$30$ were used for training the classifiers. 

\subsubsection{Classifier Selection}
\begin{figure}[!htb]
\centering
\subfloat[2 Frames]{\includegraphics[scale=0.20]{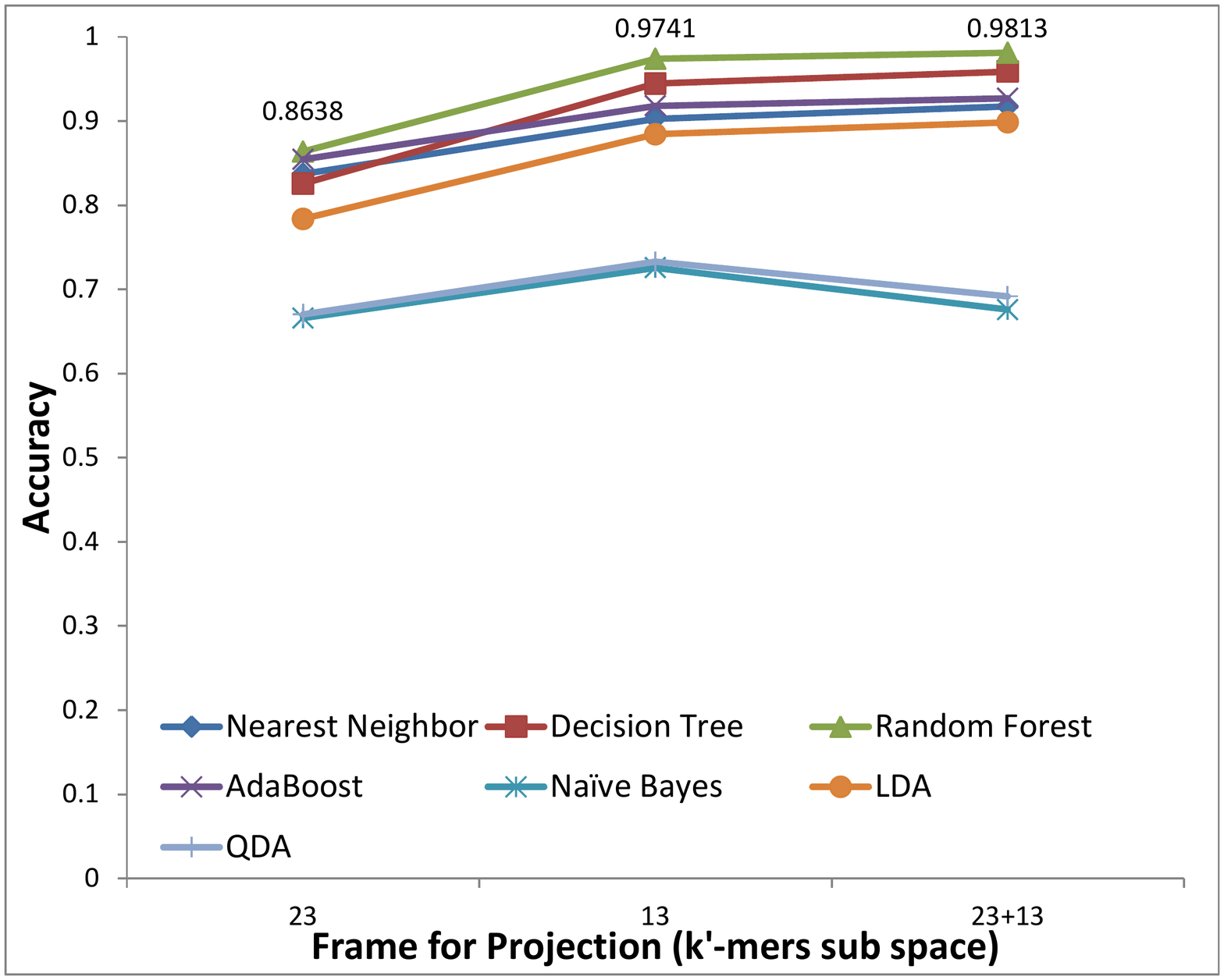}}
\qquad
\subfloat[4 Frames]{\includegraphics[scale=0.20]{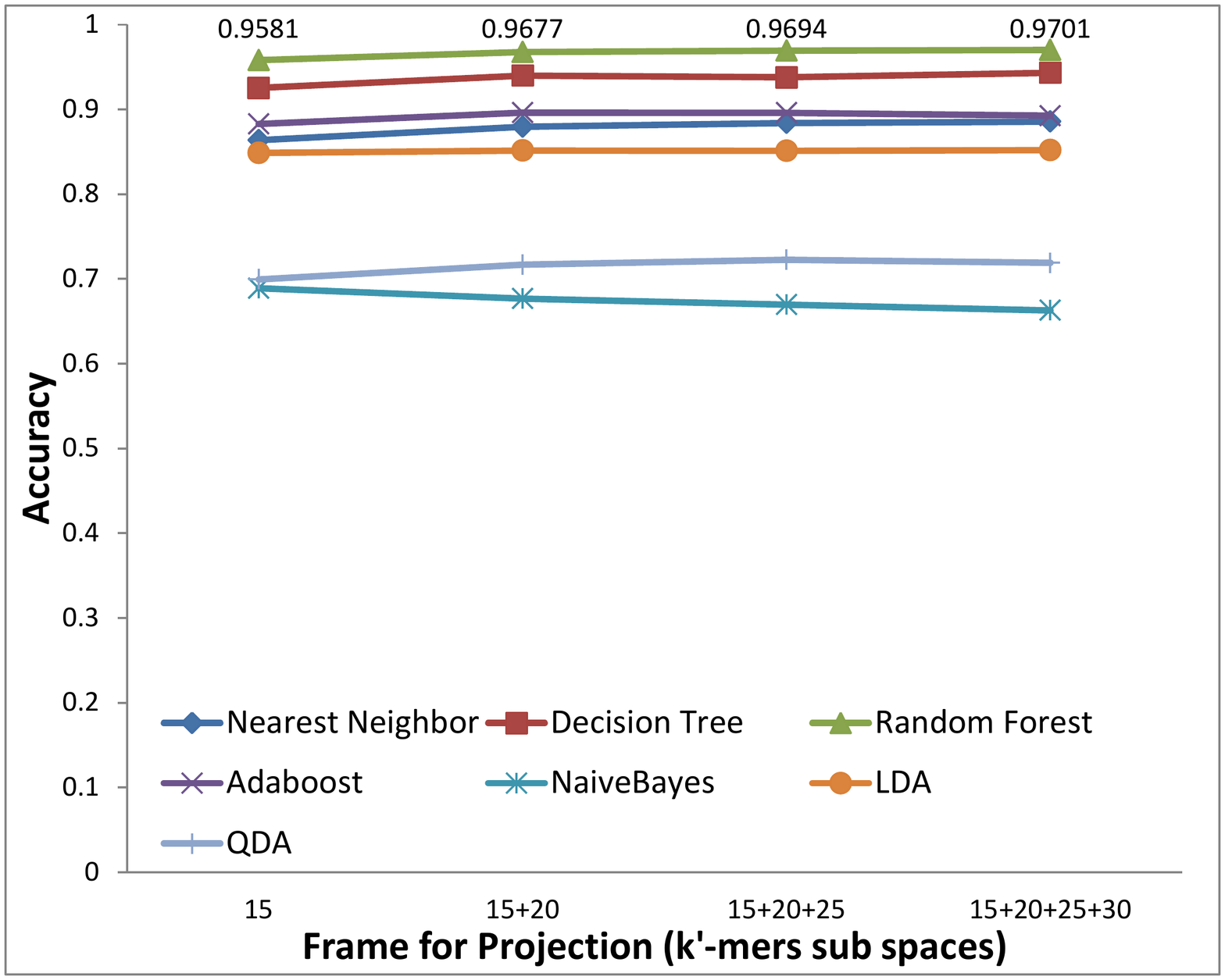}}
\caption{Accuracy of various classification algorithms for \kmer classification. $35$-mers are either projected onto a family of (a) $23$-mers, $13$-mers, and a combination of $23$-mers and $13$-mers, and (b) Combination of $(15,20,25,20)$-mers. Accuracy improves when $35$-mers are projected onto smaller sized $k^{\prime}$- mers and as the number of projections increases. Best accuracy is of the Random Forest Classifier.}
\label{fig:02}
\end{figure}

Classifiers Nearest Neighbor, Decision Tree, Random Forest, Adaboost, Naive Bayes, Linear Discriminant Analysis (LDA), and Quadratic Discriminant Analysis (QDA) are trained on the three training datasets and evaluated based on their test data accuracy over a 5-fold cross validation dataset. The classifiers are implemented in the scikit-learn library (version is 0.16.1) in python programming language (version 2.7.6). For all the classifiers, the accuracy improves as the $35$-mers are projected onto $13$-mers rather than $23$-mers (higher resolution, lower size of $k^{\prime}$-mers), and improves even further when $35$-mers are resolved onto both $13$-mers and $23$-mers (Figure \ref{fig:02}). Random Forest Classifier performs the best on all three datasets, where the accuracy for dataset (iii) is $98.12 \%$. The accuracy for Naive Bayes and QDA classifiers are lower for all datasets, and also decreases when the projections in both $13$-mers and $23$-mers are considered, indicating that inadequacy of their models for the classification of $35$-mers in these projections. The performance of other classifiers are comparable and follows similar trends. 

\vspace{-10pt}
\subsubsection{Exploring Additional Feature Spaces}
Additionally, we generate a series of 4 projections of the $35$-mers onto frames of sizes a) $\{ \mathbf{C_{15,\mathds{N}}}\}$, b) $\{ \mathbf{C_{15+20,\mathds{N}}}\}$, c) $\{ \mathbf{C_{15+20+25,\mathds{N}}}\}$, and d) $\{ \mathbf{C_{15+20+25+30,\mathds{N}}}\}$ to evaluate the effect of number of frames used for projection on the performance (Figure \ref{fig:02} b). Increasing the number of projections has no visible effect on increasing the accuracy of performance, although it increases the memory requirements and time complexity for computing counts of all five different values of $k$. Based on this, we chose the Random Forest classifier with a resolution of $35$-mers decomposed into a combination of $13$-mers and $23$-mers for other simulated and real datasets. 

\section{Results}
\subsection{Error Detection for Reconstruction of Haplotypes}
MultiRes predicts a set of error-free \kmers constituting the rare variant and conserved segments of the genomes present in a viral population. Briefly, it classifies \kmers as erroneous or rare variant \kmers based on a collection of features extracted from the sequenced reads using a Random Forest Classifier. The features for a \kmer include its count in the sequenced reads and the counts of sub-sequences (small sized \kmers) within the \kmer. The Random Forest classifier is trained on a simulated dataset where the status for a \kmer being erroneous or rare-variant is known. %

We evaluate MultiRes on simulated HIV and HCV datasets and a laboratory mixture of HIV-1 strains. MultiRes is compared to the error correction methods Quake (last checked version Feb 2012) \cite{quake}, BLESS (version 0.15) \cite{bless}, Musket (last downloaded October 2015) \cite{musket}, BFC (last downloaded October 2015) \cite{bfc}, BayesHammer (version 3.6.2) \cite{bayeshammer} and Seecer (version 0.1.3) \cite{seecer} using three measures, defined in terms of the true and erroneous \kmers. \textit{Precision} is defined as the ratio of the known true \kmers identified to the total number of \kmers predicted as true variants by an algorithm. \textit{Recall} is defined as the ratio of the true variant \kmers identified to the total number of true \kmers by an algorithm and measures the goodness of a method to retain true \kmers for a dataset. \textit{False Positives to True Positives Ratio} (FP/TP ratio) is the ratio of the erroneous \kmers predicted as true variants to the true variant \kmers identified by the algorithm. FP/TP ratio measures the number of erroneous \kmers identified by an algorithm to detect a single true variant \kmer and is a measure of the overall volume of \kmers predicted by an algorithm. The error correction method KEC \cite{kec} was not evaluated as it is applicable to 454 pyrosequencing data. ShoRAH \cite{shorah} reconstructs a set of haplotypes as a final output rather than error corrected reads and thus was not evaluated for error correction, but was used for single nucleotide variant calling and comparison.

\subsubsection{HIV Simulated Datasets}
We first assess the performance of MultiRes on the reads simulated from the HIV-1 population containing 11 haplotypes, generated from a single HIV-1 reference sequence (NC\_001802) as mentioned before. Two datasets are generated from the simulated reads: one with average haplotype coverage of 100x (denoted as HIV 100x), and second where the average coverage is 400x (denoted as HIV 400x) as increasing sequencing depth increases the absolute number of erroneous \kmers introduced in the data. We again mention that the classifier was only trained on a 10\% random sub-sample of the simulated sequenced reads and we evaluate its performance on the complete data. 

The recall of MultiRes is 95\% and 98\% on HIV 100x and HIV 400x datasets, respectively. This is comparable to the other methods where the recall is around 98\% for HIV 100x and varies from 94\% to 99\% for HIV 400x dataset (Table \ref{tab:01}). The precision of MultiRes is 89\% in the HIV 100x while the all other methods have low precisions for HIV 100x. While precision in all other methods is less than 5\% for HIV 400x dataset, the precision of MultiRes is 95\%, suggesting that precision decreases for other methods with increasing sequencing depth. Seecer and BayesHammer, methods which can handle variability in sequencing coverage, also have very low precision values compared to the proposed method.  The FP/TP ratio obtained by MultiRes are 4 to 500 times better than other methods and the number of \kmers retained is close to the true set of \kmers in the two datasets (FP/TP ratio is close to zero \& recall close to 95-98\%). Thus, while all methods retain the true \kmers to the same extent, only MultiRes reduces the number of false positive \kmers. 

\begin{table*}[!htb]
\small
\begin{center}
\caption{\bf Comparison of performance metrics of error correction methods on simulated HIV datasets \label{tab:01}}
\begin{tabular}{|c|c|c|c|c|c|c|} \hline
\multicolumn{1}{|c|}{Algorithm} & \multicolumn{2}{|c|}{FP/TP Ratio} & \multicolumn{2}{|c|}{Recall} & \multicolumn{2}{|c|}{Precision}\\ \hline
  			& HIV 100x 		& HIV 400x  		& HIV 100x 		& HIV 400x & HIV 100x	& HIV 400x \\ \hline
Uncorrected & 	53			&	121				&	98.91		& 	99.67		&	1.85 &		0.82	\\ \hline
Quake		& 	9.26		&	29.5				&	\textbf{98.63}		&	94.84		&	9.74	&	3.27 \\
BLESS		& 	0.71		&	76.7			& 	98.38		& 	99.36	&	58.48	& 	1.28	\\
Musket 		&	0.46		&	121				&	98.46		&	\textbf{99.67}	&	68.48	&	0.82	\\
BFC 		&	2.12		&	112				&	98.47		&	99.57	&	32.01	&	0.89	\\
BayesHammer &	0.37		&	69.1				&	98.47		&	98.59	&	73.04	&	1.42\\
Seecer		&	12.1		&	110				&	98.49		&	98.31	&	7.65		&	0.90\\
MultiRes 	&	\textbf{0.11}	&	\textbf{0.048}	&	95.01	&	98.17	&	\textbf{89.34}	&	\textbf{95.39}	\\
\hline
\end{tabular}
\end{center}

The False positive/True Positive ratios (FP/TP ratios), Recall, and Precision are compared on two HIV datasets for the methods: Quake, BLESS, Musket, BFC, BayesHammer, Seecer, and the proposed method MultiRes. The error corrected reads from each method are broken into \kmers and compared to the true \kmers in the HIV-1 viral populations. Uncorrected denotes the statistics when no error correction is performed. Bold in each column indicates the best method for the dataset and the metric evaluated. 
\end{table*}

\subsubsection{Robustness: Testing MultiRes on a Hepatitis C Virus Dataset}
We also evaluate our method on reads simulated from viral populations consisting of the E1/E2 gene of Hepatitis C virus (HCV). The purpose of using HCV strains is to understand the performance of MultiRes for detecting sequencing errors and rare variants when the complexity of the dataset varies significantly from the training data used for MultiRes. Two HCV populations observed in patients in previous studies are used as simulated viral populations. The first, denoted as HCV 1, consists of 36 HCV strains from E1/E2 region and are of length 1672 bps \cite{hussein2014new}. The second, denoted as HCV 2, consists of 44 HCV strains from the E1/E2 regions of the HCV genome with lengths 1734 bps \cite{vga,kec}. We simulate 500K Illumina paired end reads from both datasets under a power law (with ratio 2) of reads distribution amongst the strains \cite{grinder}. The two simulated datasets are denoted as  HCV1P and HCV2P respectively. The power law distribution of reads also helps in evaluating the performance of MultiRes when more than 50\% of the haplotypes are present at less than 5\% relative abundances.

\begin{table*}[!htb]
\small
\begin{center}
\caption{\bf Comparison of performance metrics of different methods on HCV population datasets\label{tab:02}}
\begin{tabular}{|c|cc|cc|cc|} \hline
\multicolumn{1}{|c|}{Algorithm} & \multicolumn{2}{|c|}{FP/TP Ratio} & \multicolumn{2}{|c|}{Recall} & \multicolumn{2}{|c|}{Precision}\\ \hline

{} 			& {HCV1P} &  {HCV2P}  & {HCV1P} & {HCV2P} & {HCV1P} & {HCV2P} \\ \hline
Uncorrected	&	1201	&	571		&	99.51	&	99.88		&	0.08	& 0.17\\ \hline
Quake		&	303.3	&	149		&	96.41	&	97.23	&	0.32	&	0.66\\
BLESS 		&	202		&	112		&	98.35	&	97.18		&	0.49	& 0.88	\\
Musket		&	938		&	463		&	93.53	&	89.17		&	0.10	&	0.21\\
BFC			&	352		&	161		&	99.32		&	99.84			&	0.28	&	0.61\\
BayesHammer	&699	&	340		&	98.12		&	97.1		&	0.14	&	0.29\\
Seecer			&1095	&	528		&	\textbf{99.48}&\textbf{99.85}	&0.09	&	0.19\\ 
MultiRes		&	\textbf{37.4}	&	\textbf{19.54}	&	96.5	&	94.25	&	\textbf{2.6} &	\textbf{4.87}\\
\hline
\end{tabular}
\end{center}

The false positive to true positive ratios, recall, and precision of error correction methods on the two simulated HCV datasets are shown. Uncorrected refers to the statistics when no error correction is performed. Bold font in each column indicates the best method for each dataset on the evaluated measure. 
\end{table*}

All methods have recall greater than 90\% on both datasets (Table \ref{tab:02}). Again, the difference between MultiRes and other methods is evident from the FP/TP ratios and precision. The false positive to true positive ratios for MultiRes are less than other methods at least by a factor of 5 (Table \ref{tab:02}). MultiRes still outperforms all other methods on predicting the smallest set of predicted \kmers while maintaining high recall levels of true \kmers. 

The recall for MultiRes is respectively 96\% and 94\% on HCV1P and HCV2P datasets, which is less than the method Seecer that has recall values around  99\%. Seecer marks more than 90\% of the observed \kmers as true, which explains the high recall values. However, this also leads to a large number of false positive \kmers being predicted as true \kmers in Seecer, leading to low precision values. All other methods also achieve high recall by retention of all large fraction of observed \kmers, as indicated by their precision values being less than 1\% and false positive to true positive ratios being greater than 100. 

The similar performance of MultiRes on a dataset, such as the HCV population, which is diverse in genome composition from the simulated HIV-1 sequences used in simulation indicates that robustness of the Random Forest Classifier in MultiRes. The classifier is capturing properties of the Illumina sequencing platform and the fact that both datasets contain a large number of rare variants occurring at \kmer counts close to the sequencing errors. Thus, MultiRes can be used as it is for error and rare variant detection in diverse datasets.

\subsubsection{Performance of MultiRes with variation in counts of \kmers} 
\begin{figure}[!htb]
\centering
\subfloat[Total \kmer Multiplicity Plots]{\includegraphics[scale=0.25]{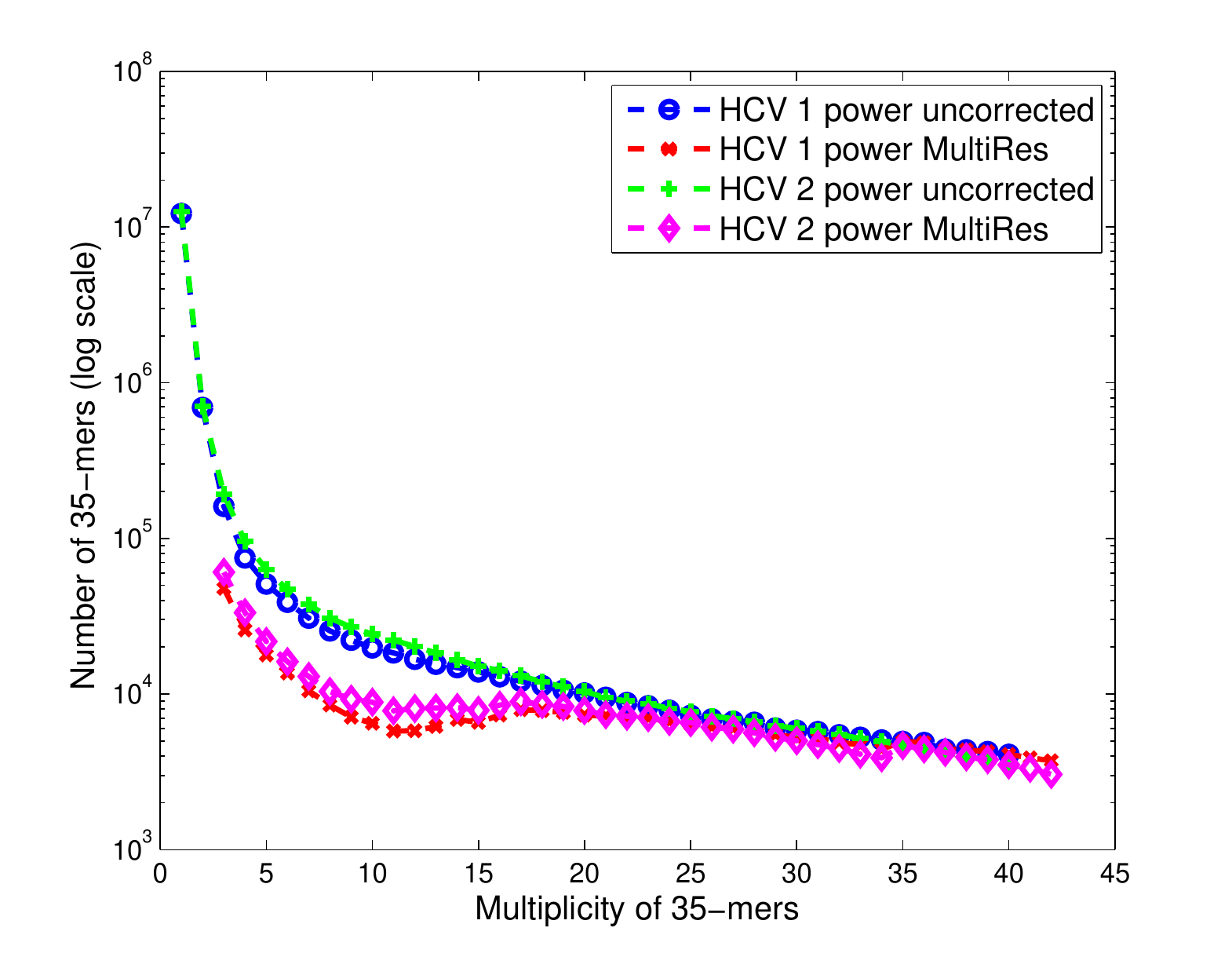}} 
\qquad
\subfloat[True rare Variants Multiplicity plots]{\includegraphics[scale=0.25]{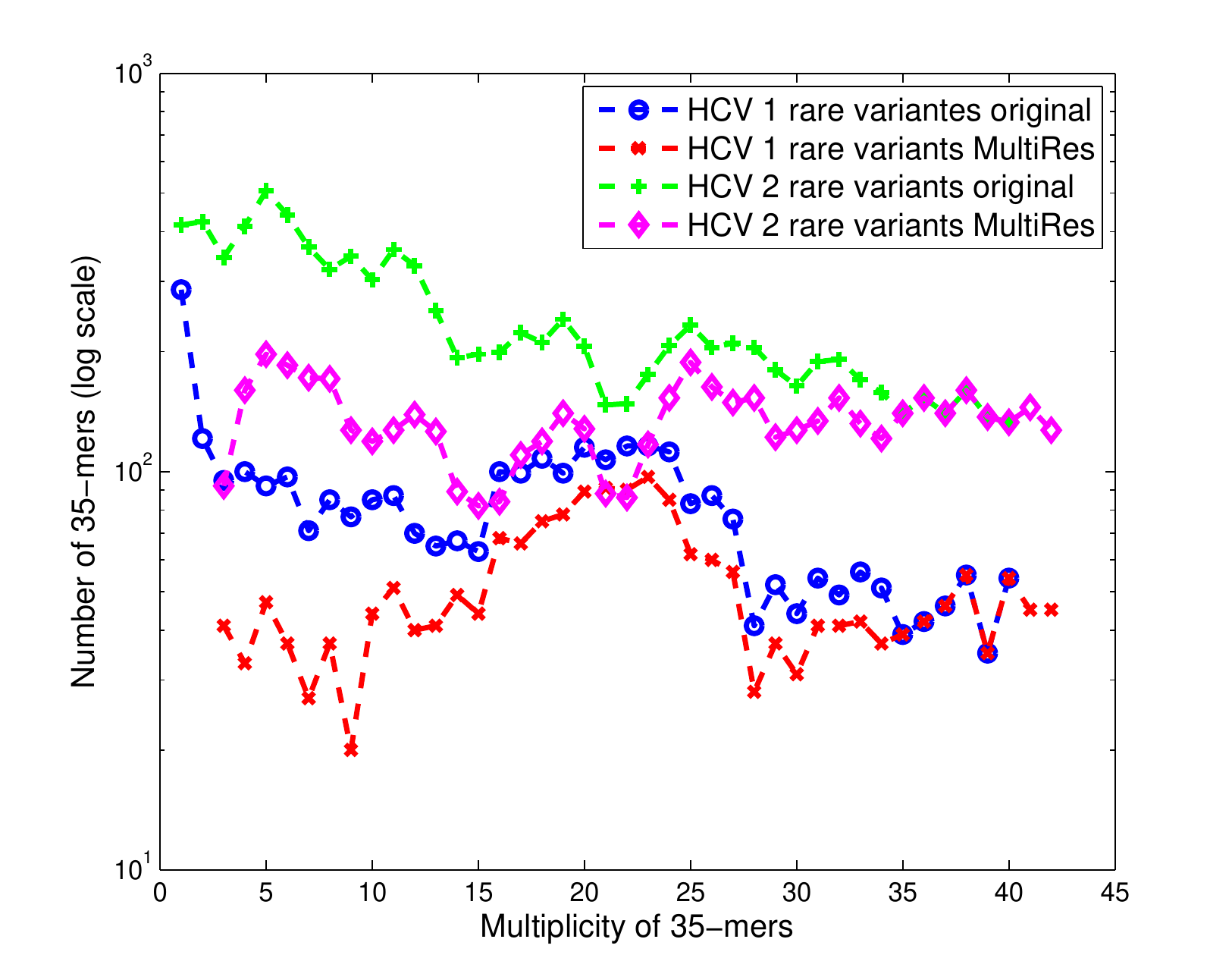}}
\caption{35-mer multiplicity plots for HCV1P and HCV2P datasets denoting (a) The predicted true $35$-mers from MultiRes (HCV1P red, HCV2P pink) compared to the uncorrected data (HCV1P blue,HCV2P green), and (b) The true positive rare variants $35$-mers from MultiRes (HCV1P red, HCV2P pink) versus the ground truth $35$-mers (HCV1P red, HCV2P pink).}
\label{fig:05}
\end{figure}

We investigate the performance of MultiRes to distinguish a \kmer as erroneous or rare variant as the count of the sequenced \kmer varies. MultiRes predicts about one-fourth of the observed \kmers as rare variants for \kmer counts less than 15, and predicts almost all of the observed \kmers as true for counts greater 20 (Figure \ref{fig:05} (a)). This suggests that MultiRes predicts rare variant \kmers for all observed counts and detects more rare variant \kmers than a method based on a single threshold. Most of the \kmer based error correction methods use a single threshold over the \kmer counts, which will clearly lose true rare variant \kmers (Figure \ref{fig:05} (b)). On the other hand, MultiRes has a recall of 50\% for \kmers observed 3 times, while still correctly identifying more than 75\% of the \kmers as erroneous. The recall of MultiRes increases to 100\% as the counts of the observed \kmers increases to 35. This indicates the importance of not having a single threshold for distinguishing between sequencing errors and rare variants in viral population datasets, and our MultiRes bypasses a single threshold by training a Random Forest classifier. 


\subsubsection{Evaluation on Population of 5 HIV-1 Sequences}
We also evaluate MultiRes on a laboratory mixture of five known HIV-1 strains \cite{fiveviralmix}, which captures the variability occurring during sample preparation, errors introduced in a real sequencing project, and mutations occurring during reverse transcription of RNA samples. Five HIV-1 strains (named YU2, HXB2, NL43, 89.6, and JRCSF) of lengths 9.1 kb were pooled and sequenced using Illumina paired end sequencing technology (Refer to \cite{fiveviralmix} for details). Each HIV strain was also sequenced separately in their study and aligned to their known reference sequence (from  Genbank) to generate a consensus sequence for each HIV-1 strain \cite{fiveviralmix}.  This provides us with a dataset of actual sequence reads where the ground truth is known allowing us to assess the performance of MultiRes and other methods. We extracted $35$-mers from the paired end sequencing data and classify them using the Random Forest classifier of MultiRes trained on the simulated HIV sequencing data. 

All the error correction methods including MultiRes have recall values around 97\%, indicating that the performance for recovery of true \kmers is comparable across all methods. The false positive to true positive ratio for MultiRes is 13 while all other methods have ratios more than 120. MultiRes predicts 359 thousand unique \kmers in the set of true \kmers while all other methods predict more than 5 million unique \kmers. Even methods that take variance in sequencing depths while performing error correction, such as BayesHammer and Seecer, predict 11.3 million and 6.3 million unique \kmers which is two orders more than the ground truth number of \kmers in the consensus sequence of the 5 HIV-1 strains (53 thousand unique \kmers). Thus, even considering the artifacts introduced in sequencing, MultiRes has by far the most compact set of predicted error free \kmers amongst all methods while retaining high number of true \kmers. 

\subsubsection{Runtime and Memory}
\begin{figure}
\centering
\subfloat[Example 7-mer]{\includegraphics[scale=0.25]{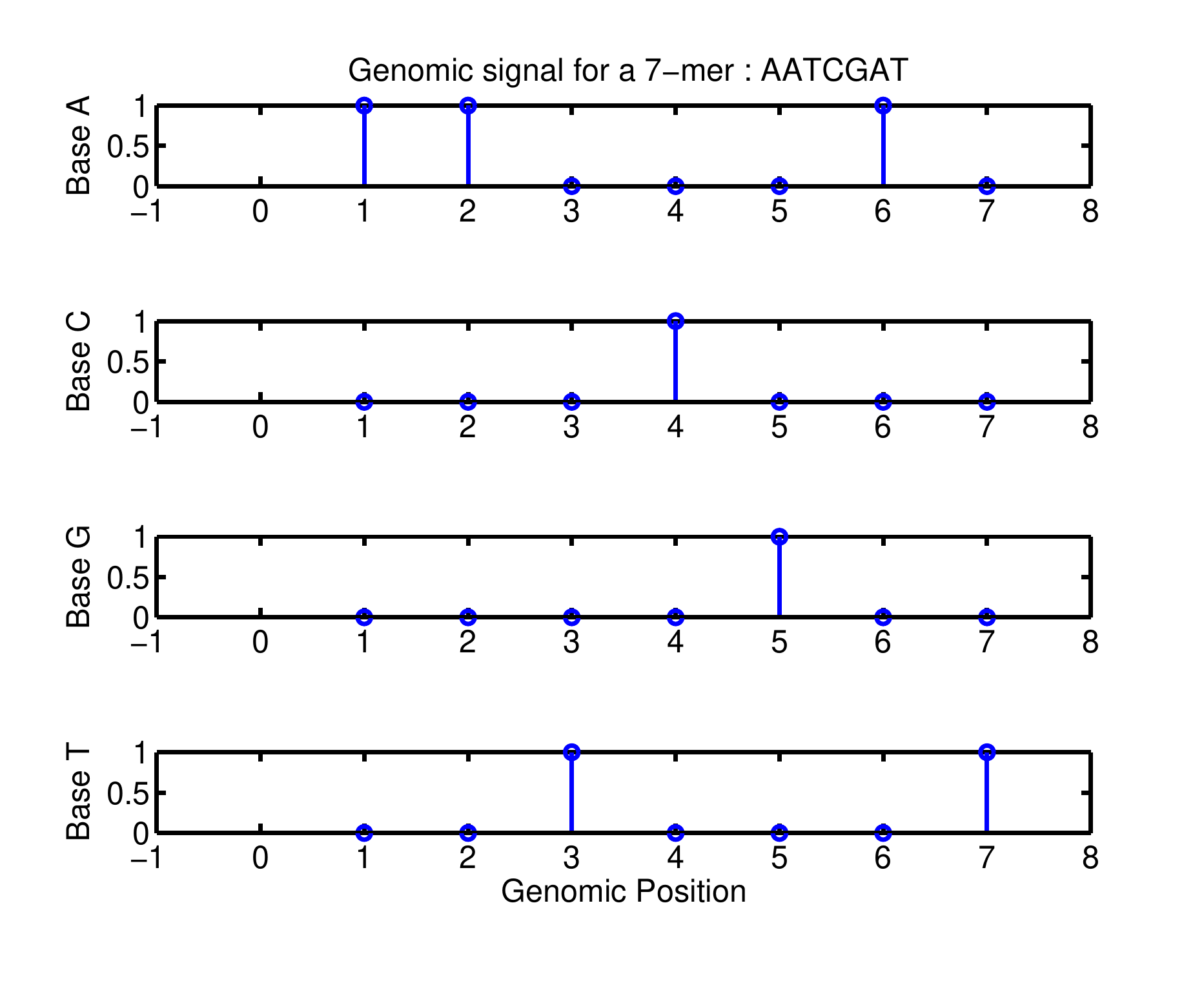}}
\qquad
\subfloat[Timings information]{
\includegraphics[scale=0.25]{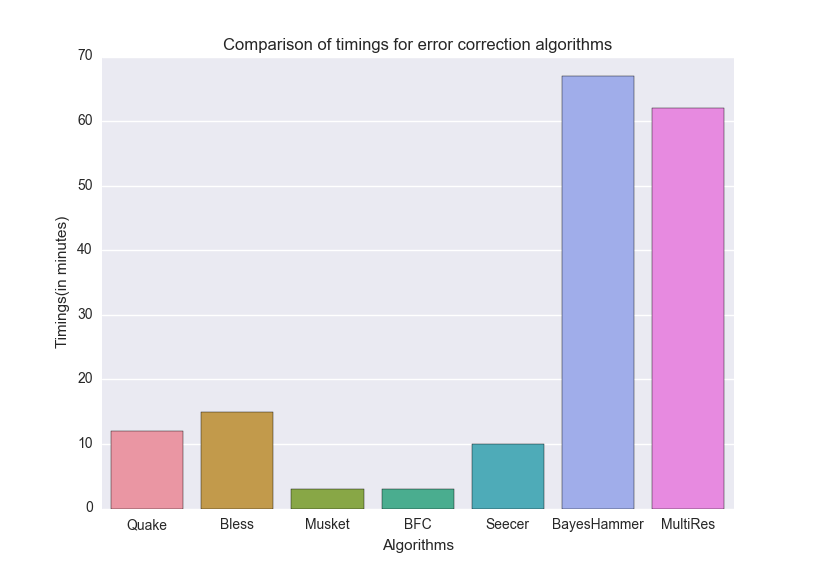}}
\caption{(a) A genomic signal example for a 7-mer AATCGAT that belongs to the set $\mathbf{C_7}$. (b) Comparison of running times for different algorithms on 5-viral mix dataset on 8GB memory nodes of 2X Dual Core AMD Opteron 2216 systems from Dell. The time noted for BayesHammer is only the time reported for BayesHammer error correction step in spades (version 3.6.2). The time reported for MultiRes is the combined time for \kmer counting, predicting \kmers as erroneous and rare variants and generating the final output. \label{fig:Timings}}
\end{figure}

MultiRes has comparable running times to BayesHammer on the five-viral mix dataset (Figure \ref{fig:Timings}) on a Dell system with 8GB main memory, and 2X Dual Core AMD Opteron 2216 CPU type. The performance on all other datasets was similar indicating that the timings are comparable. Additionally, while other methods have parallel implementations, the error correction classifier step in MultiRes is a single thread serial implementation. Thus, its runtime can be significantly improved via parallelization of \kmer prediction step. 

\begin{table*}[!htb]
\small
\begin{center}
\caption{\bf Comparison of performance metrics on 5-viral mix HIV-1 dataset. \label{tab:03}}
\begin{tabular}{|c|c|c|c|} \hline
\MyHead{2cm}{Algorithm}	&	\MyHead{2cm}{Recall}	&	\MyHead{2cm}{FP/TP Ratio}	& \# of unique $35$-mers \\ \hline
Uncorrected	&	98.01	&	439	& 11.4 M \\
BLESS 	&	97.31	&	227	&	5.89 M \\
Musket 	&	\textbf{97.91}	&	366	& 11.2 M \\
BFC 	&		97.55		&	316	& 9.6 M\\
BayesHammer & 97.49		&	122 &	6.3M\\
Seecer	&	97.84			& 	220	&	11.3M\\
MultiRes 	&	96.64	& \textbf{13}	& \textbf{359 K} \\
\hline
\end{tabular}
\end{center}

The recall and FP/TP Ratios of each method are evaluated on the 5-viral mix HIV-1 dataset. The number of unique $35$-mers indicates the number of unique $35$-mers predicted by a method. There are 53 thousand true unique $35$-mers in the consensus sequences of the 5 viral strains. Bold indicates the best method for the measure in each column.
\end{table*}

\subsection{Comparison of MultiRes to Variant Calling Methods for Viral Populations}
As one of the objectives in NGS studies of viral populations is to identify the single nucleotide polymorphisms (SNPs) in a population \cite{reviewpaper,shorah,vphaser2} which is sensitive to erroneous reads, we evaluate the inference of SNPs from the \kmers predicted by MultiRes, and compare it to known SNP profiling methods for viral populations. We first align the predicted \kmers from MultiRes to a reference sequence of the viral population and a base is called as a SNP when its relative fraction amongst the \kmers aligned at that position is greater than $0.01$. We use the command in samtools: \textit{samtools mpileup} \cite{samtools1,samtools2} to determine the number of \kmers aligned at a position on the reference sequence. All the variants that occur at a frequency greater than the error threshold at that position are reported as SNPs. The choice of the reference sequence is based on the viral population data being evaluated, and the same reference sequence is used for calling true SNPs and the SNPs predicted by a method. 

Each SNP detected at a base position of the reference and detection of the reference base itself are treated as \textit{true positives} for a method; thus the number of true positives can be greater than the length of the reference sequence. All the SNPs predicted by a method and the number of bases mapped to the reference sequence are known as the \textit{total SNP predictions} of a method. We use three measures for evaluating the SNPs called by any method. \textit{Precision} is defined as the ratio of the number of \textit{true positives} to the \textit{total SNP predictions} made by a method, while \textit{recall} is defined as the ratio of the \textit{true positives} to the total number of SNPs and reference bases in the viral population. Finally, \textit{false positive to true positive ratio} is a ratio of the number of false SNP predictions to the number of \textit{true positives} detected by a method. 

We compare our results to state-of-the-art variant calling methods for viral populations VPhaser-2 \cite{vphaser2}, a rare variant calling method LoFreq \cite{lofreq}, and viral haplotype reconstruction algorithm ShoRAH \cite{shorah} using the above three measures. The reference sequence used by variant calling methods VPhaser-2 and LoFreq is the same as that used by \textit{samtools} to determine the true SNPs, while the SNPs predicted by ShoRAH at default parameters are compared directly to the true SNPs. We only used the SNP calls from VPhaser-2 for evaluation, as length polymorphisms are not generated by the other methods, but the results from VPhaser-2 were not penalized when comparing the SNPs. 

We report results for LoFreq \cite{lofreq}, VPhaser \cite{vphaser2}, ShoRAH \cite{shorah} and our method MultiRes on all datasets (Table \ref{tab:04}). Overall, MultiRes has greater than 94\% recall and precision values greater than 83\% in all datasets. LoFreq and VPhaser have comparable recall but lower precision values and an increase in the FP/TP ratios on the HCV population datasets, indicating a decrease in performance. ShoRAH overall has lower recall values, nevertheless a 100\% precision in all but the 5-viral mix dataset, suggesting that it misses true SNPs but is very accurate when it calls a base as SNP. Overall all methods have low values for FP/TP ratio as compared to before,  indicating that the number of false positive SNP predictions are low. The metric where MultiRes outperforms others is the lowest number of true SNPs missed. This shows that even with a simplistic SNP prediction method used in MultiRes, it is able to capture the true variation of the sampled viral population and has the lowest false negatives of well established methods. This demonstrates that using error-free set of \kmers can vastly increase the variant detection in viral populations. 

The number of reads or \kmers aligned to the reference sequence are comparable across the methods, except for HCV2P dataset where MultiRes has 85\% \kmers mapped compared to 100\% read mapping (Table \ref{tab:04}). It is possible that the unmapped \kmers correspond to the length variants and could be verified by haplotype reconstruction using the predicted \kmers, but that was not the focus in this paper.

\begin{table*}
\small
\begin{center}
\caption{\bf Comparison with Variant Calling methods on all datasets \label{tab:04}}
\begin{tabular}{|c|c|c|c|c|c|c|} \hline
    Dataset & 	\MyHead{2cm}{Method}& \MyHead{1.5cm}{Recall (\%)} 	&\MyHead{1.5cm}{FP/TP Ratio} 	& \MyHead{1.5cm}{Precision (\%)} 		& \# of False Negatives & Mapped Reads (\%)\\ \hline
	&			LoFreq		&	97.33		&	0.004		&	{99.60}		&	444 &	89.51\\
HIV 100x	&	Vphaser		&	98.90		&	0.007		&	99.26				&	183	&	89.51\\
			& 	ShoRAH		&	55.21		&	\textbf{0}	&	\textbf{100}			&	7746	&	\textbf{98.04}\\
	&			MultiRes	&	\textbf{99.69}	&	0.011	&	98.88			&	\textbf{51}	& 97.89\\ \hline
	
	&			LoFreq		&	84.83		&	0&	99.99		&	2522	& 99.55\\
HIV 400x	&	Vphaser		&	\textbf{95.92}	&	0.292	&	77.37			&	\textbf{678} & 99.55\\
			& 	ShoRAH		&	55.21		&	\textbf{0}			&	\textbf{100}		& 		7746	& \textbf{99.95}	\\
	&			MultiRes   &	95.57		&	0.007		&	99.33				&	736	& 97.34\\ \hline
	
	&			LoFreq		&	\textbf{98.30}	&	1.282	&	43.82			&	\textbf{31}	& \textbf{99.99}\\
HCV1P	&	Vphaser		&	93.51		&	1.628		&	38.05				&	118	& \textbf{99.99}\\
			& 	ShoRAH		&91.92		&	\textbf{0}			& 	\textbf{100}			&	147	&	\textbf{99.99}\\
	&			MultiRes	&	98.24		&	0.597&	62.64		&	32	& 97.32\\ \hline
	
    &			LoFreq		&	97.10		&	1.046		&	48.87		&	60 & \textbf{100}	\\
HCV2P	&	Vphaser		&	95.65		&	1.492		&	40.13			&	90  & \textbf{100}	\\
			& 	ShoRAH		&83.73		& 	\textbf{0}			&	\textbf{100}		&	337	&	99.95	\\
	&			MultiRes	&	\textbf{98.79}	&	0.201&	83.27		&	\textbf{25}	& 85.14 \\ \hline
	
	&			LoFreq		&	99.06		&	0.085		&	92.15				&	101	& 98.59\\
5-viral mix	&	Vphaser		&	92.68		&	0.039		&	96.25		&	789	& 98.59 \\
			& 	ShoRAH		& 98.66			&	\textbf{0.014}		&	\textbf{98.99}		&	109 &	\textbf{99.3}\\
	&			MultiRes	&	\textbf{99.39}	&	0.077	&	92.82				&	\textbf{66}	& 96.29 \\ \hline
\end{tabular}
\end{center}

The recall, false positive to true positive ratios (FP/TP), precision, number of false negatives, and \% of mapped reads by methods LoFreq, VPhaser-2, ShoRAH, and MultiRes are computed for listed datasets. Outputs from LoFreq (version 2.1.2), VPhaser-2 (last downloaded version October 2015), and ShoRAH (last downloaded version from November 2013) are compared against known variants for simulated datasets. For 5-viral mix, the consensus reference provided by \cite{fiveviralmix} was used to determine ground truth variants. MultiRes variants are determined by aligning $35$-mers to a reference sequence and bases occurring at more than $0.01$ frequency as variants. Bold for each dataset indicates the best method for the performance measures.
\end{table*}

\section{Discussion and Conclusions}

We have proposed a frame-based representation for genomes and developed the classifier MultiRes for identifying  rare variant and erroneous \kmers obtained from Illumina sequencing of viral populations. Our method does not rely on a reference sequence and uses concepts from signal processing to justify using the counts of a set of \kmers for the classification of a \kmer. We first gave a representation of haplotype genomes as four-dimensional spatial signals and demonstrate that \kmers of a fixed size form a mathematical frame for the genomic signals. Next, we showed that the projection of the sampled reads signals are maximized onto the signals corresponding to its constituent \kmers and utilize the projections of sampled reads signals onto multiple frames as features for our classifier MultiRes. 

We demonstrated the performance of MultiRes on simulated HIV and HCV viral populations and real HIV viral populations containing viral haplotypes at varying relative frequencies, where it outperformed existing error correction methods in terms of recall and the total number of predicted \kmers. Though most error correction methods evaluated assumed that sequenced reads originated from a single genome sequenced at uniform coverage, our method also works better than the method BayesHammer, which can tackle non-uniform sequencing coverage, and the method Seecer, which can additionally deal with alternative splicing and polymorphisms. 


The error corrected \kmers predicted by MultiRes enable the usage of \textit{de novo} assembly methods for closely related genomes. A major challenge for using De Bruijn graph based methods for viral populations has been the increased complexity of the graph due to the presence of large number of sequencing errors \cite{vicuna}. The output of our method is a set of predicted error free \kmers that can be directly used as an input to \textit{de novo} De Bruijn graph based assembly methods \cite{spades,cortex} for reconstruction of viral haplotypes in the population, and for calling structural variants directly from the assembly graph. MultiRes has high recall of true \kmers while outputting the least number of false positive \kmers, thereby making \textit{de novo} assembly graphs manageable.  

Another end-goal of studying viral populations is to understand the single nucleotide variation in the viral population sample. We also tested MultiRes for prediction of SNPs and showed that the error corrected \kmers can be directly used for SNP calling. The SNPs called by MultiRes' data has either the highest or the second highest recall of the SNPs compared to other methods for viral population, demonstrating its comparable performance for such analysis. 

MultiRes relies on the counts of multiple sizes of \kmers observed in the sequenced reads, and the choice of \kmer length is an important parameter. The minimum value of $k$ chosen should be such that a \kmer can only be sampled from a single location in the genome. This is possible in viral populations where there are small repeats present. Choosing the number of \kmer sizes used is another parameter, and while accuracy can be improved by increasing it, additional \kmer counting increased the number of computations. As demonstrated by our experiments, choosing three different values of $k$, namely ($k, 2\cdot k, 3\cdot k$) was sufficient for accurate results. 

MultiRes was primarily developed for detection of sequencing errors and rare variants in viral populations, which have small genomes. Extending our method for larger genomes may require additional tuning of the parameters via re-training of the classifier, but the concepts developed here are applicable to studying variation in closely related genomes such as cancer cell lines. It is also applicable for understanding somatic variation in sequences as their variation frequency is close to the sequencing error rates. The technique can also be explored for newer sequencing machines, such as PacBio sequences and Oxford Nanopore long read sequencing, where the type of sequencing errors are different, but the concepts of projections of signals are still applicable. The software is available for download from the github link (\url{https://github.com/raunaq-m/MultiRes}).


\bibliographystyle{plain}

\end{document}